\documentclass[11pt,a4paper]{article}

\usepackage{amsmath,amsfonts,amssymb,amsbsy,textcomp}
\usepackage{graphicx,color}
\usepackage{hyperref}
\usepackage{graphics}

\usepackage{amsmath, amssymb,graphicx}
\usepackage{epsfig}
\usepackage{verbatim}
\usepackage{amsfonts}

\usepackage{epsf}
\def\ssl#1{\rlap{\hbox{$\mskip 3 mu /$}}#1}
\def\be{\begin{equation}}
\def\ee{\end{equation}}
\def\bc{\begin{center}}
\def\ec{\end{center}}
\def\ba{{\bar{a}}}
\def\bR{{\mathbb{R}}}
\def\bD{{\mathbb{D}}}

\def\by{{\mathbf{y}}}
\def\bbZ{{\mathbb{Z}}}
\def\cB{{\mathcal{B}}}
\def\cD{{\mathcal{D}}}
\def\cG{{\mathcal{G}}}
\def\cF{{\mathcal{F}}}
\def\cM{{\mathcal{M}}}
\def\cN{{\mathcal{N}}}

\def\cR{{\mathcal{R}}}

\def\cT{{\mathcal{T}}}
\def\cI{{\mathcal{I}}}
\def\cJ{{\mathcal{J}}}

\def\cK{{\mathcal{K}}}
\def\cX{{\mathcal{X}}}

\def\ga{\gamma}
\def\gae{n_e}
\def\gam{n_m}

\def\sf{{\rm sf}}

\def\r2{{\sqrt{2}}}

\def\z{{\zeta}}
\def\bZ{\bar{Z}}

\def\th{{\theta}}

\def\cR{{\mathcal{R}}}
\def\w{\omega}
\def\Tr{{\rm Tr}}

\def\eff{{\rm eff}}

\begin{document}

\begin{titlepage}

\begin{flushright}
{\bf \today} \\
MAD-TH-10-01\\
\end{flushright}
\begin{centering}
\vspace{.2in}

{\Large {\bf Wall Crossing and Instantons in Compactified Gauge Theory}}

\vspace{.3in}

Heng-Yu Chen${}^{1}$, Nick Dorey${}^{2}$ and Kirill Petunin${}^{2}$\\
\vspace{.2 in}
${}^{1}$Department of Physics, University of Wisconsin, \\
Madison, WI 53706, USA.\\
\vspace{.2in}
and \\
\vspace{.1 in}
${}^{2}$DAMTP, Centre for Mathematical Sciences \\
University of Cambridge, Wilberforce Road \\
Cambridge CB3 0WA, UK \\
\vspace{.2in}
{\bf Abstract}
\\
\end{centering}
We calculate the leading weak-coupling
instanton contribution to the moduli-space metric of
${\cal N}=2$ supersymmetric Yang-Mills theory with gauge group $SU(2)$
compactified on $\mathbb{R}^{3}\times S^{1}$. The results are in
precise agreement with the semiclassical expansion of the exact metric
recently conjectured by Gaiotto, Moore and Neitzke based on considerations
related to wall-crossing in the corresponding four-dimensional theory.
\vspace{.1in}

\end{titlepage}

\paragraph{}

\section{Introduction}
\paragraph{}
Supersymmetric gauge theories provide a setting
where interesting non-perturbative phenomena such as confinement and chiral
symmetry breaking can be described with exact
analytic formulae \cite{SW1,SW2}. Results of this kind rely on the powerful
constraints of supersymmetry and also on conjectured duality
properties of the theories in question. Often the resulting exact
formulae can be expanded at weak coupling where they provide a
precise prediction for instanton effects which can then be compared to
direct semiclassical computations in weakly coupled gauge theory \cite{DHKM}.
Such calculations then provide a strong check on conjectured
dualities and other assumptions underlying the exact analysis.
\paragraph{}
For the case of ${\cal N}=2$ supersymmetric Yang-Mills theory in four
dimensions, an exact description of the low-energy effective action
and of the BPS spectrum was provided by Seiberg and Witten (SW) in
\cite{SW1}\footnote{Direct semiclassical tests of the Seiberg-Witten
solution itself were intiated in \cite{FP,DKM} and eventually completed
in \cite{Nek}. }. More precisely, for a given value of the electric
and magnetic charges, the SW solution determines the mass of the
corresponding BPS state if it is present in the theory. However,
except in the simplest cases, the
question of which states are present in the spectrum at a given
point of the moduli space remained open. In the weak coupling regime
the question can be answered by a semiclassical analysis of the
spectrum of monopoles and dyons. However, the moduli space contains
walls of real codimension one on which BPS states become marginally
stable and can decay. Recently Kontsevich and Soibelman
\cite{KS1,KS2} proposed a
precise algorithm which determines how the spectrum changes across
these curves of marginal stability in the moduli space. Given the weak
coupling spectrum, their results effectively determine the
spectrum at any point in the moduli space.
\paragraph{}
Another perspective on the wall-crossing conjecture of \cite{KS1,KS2}
is obtained by considering the behaviour of the corresponding theory
compactified down to three dimensions on a circle. After
compactification, the BPS states of the four-dimensional theory yield
distinct instanton\footnote{Here the instantons refered to are really
instantons of the low-energy effective theory including all quantum
corrections. They should be distinguished from the
instantons appearing in the semiclassical
calculation presented below
which correspond to finite-action solutions of the equations of motion of
the underlying gauge theory.} corrections to the hyper-K\"{a}hler metric on the
moduli space of the compactified theory. On general grounds this
metric is expected to be smooth and at first sight this seems to be in
contradiction with the discontinuous changes in the four-dimensional
BPS spectrum across walls of marginal stability. A remarkable
resolution of this puzzle has recently been suggested by Gaiotto,
Neitzke and Moore (GMN) \cite{GMN1}. They showed that the a smooth metric
can be obtained precisely when the spectrum obeys the Kontsevich-Soibelman
wall crossing formula. Indeed, this condition leads to a set of
integral equations which determine the moduli-space metric exactly for
any value of the compactification radius! The main purpose of this
paper is to investigate the weak-coupling contents of their results,
and to test these against first-principles semiclassical calculations.
\paragraph{}
In the following we will focus on the simplest case, of ${\cal N}=2$
SUSY Yang-Mills theory with gauge group $SU(2)$ compactified to three
dimensions on circle of radius $R$. The moduli space is parameterised
by a single scalar VEV, $a$, and the dynamical scale of the
corresponding four-dimensional theory is denoted $\Lambda$.
In the weak-coupling limit, $|a|\gg |\Lambda|$
we will see that the GMN results give rise
to a sum over semi-classical instanton contributions carrying magnetic
charge. Each instanton contribution has
a prefactor which is a complicated function of dimensionless parameter
$|a|R$. As found in
previous investigations of instanton effects in compactified gauge
theory \cite{Dorey2000A,Dorey2000B}, the leading semiclassical
contribution can be expanded in two distinct ways. For $|a|R\gg 1$ the
result can be expanded as a sum over the contributions of magnetic
monopoles and Julia-Zee dyons regarded as classical solutions of
finite Euclidean action on $\mathbb{R}^{3}\times S^{1}$. We focus on
the contributions of unit magnetic charge but arbitrary electric
charge. We reproduce these terms in the GMN result by a direct
semiclassical calculation. An important feature first noticed in
\cite{Dorey1997} is that, unlike similar calculations in four
dimensions, the functional determinants corresponding to fluctuations
of the bose and fermi fields do not cancel. We evaluate the ratio of
fluctuation determinants from first principles
and find that it precisely reproduces the prefactors appearing in the
weak-coupling expansion of the GMN results. Given the relation between
the GMN integral equations and the Kontsevich-Soibelman conjecture,
this can also be regarded as a first-principles test
of the latter \footnote{Strictly speaking the full spectrum is already
known in this simple case and agrees with the conjecture of
\cite{KS1,KS2}. However the resulting formula of \cite{GMN1}
for the hyper-K\"{a}hler
metric on the moduli space of the compactified theory still yields
non-trivial predictions at weak coupling. In addition, we expect
our calculation to generalise straightforwardly to cases with larger
gauge groups and/or additional matter where the wall-crossing formula
remains conjectural.}.
\paragraph{}
As one goes towards smaller values of the dimensionless radius $|a|R$,
the sum over the electric charges of the semiclassical dyons diverges
and requires Poisson resummation. As in
\cite{Dorey2000A,Dorey2000B}, the resulting series also admits an
interpretation in terms of classical configurations of finite
action. The relevant configurations are
an infinite tower of twisted monopoles obtained by applying large
gauge transformations to the BPS monopole \cite{LeeYi}. Finally, in
the three-dimensional limit, $|a|R\rightarrow 0$, the twisted monopoles
decouple and only the contribution of a single BPS monopole remains.
In the context of ${\cal N}=4$ SUSY Yang-Mills realised on the world
volume of two parallel D3 branes in IIB string theory,
the relation between the two
corresponding expansions can be understood as T-duality to an
equivalent configuration of D2 branes in the IIA theory. It would be
interesting to find a similar stringy interpretation in the present
context.
\paragraph{}
Another interesting question concerns the three-dimensional limit of
the GMN results. Taking the limit $|a|R\rightarrow 0$ with the effective
three dimensional coupling held fixed, we obtain ${\cal N}=4$ SUSY
Yang-Mills in three dimensions with gauge group $SU(2)$.
The exact metric on the moduli space of this theory was conjectured to
coincide with the Atiyah-Hitchin metric in \cite{SW3D}. This proposal was
then tested against an explicit semiclassical calculation of the one-monopole
contribution \cite{Dorey1997}. Taking the three-dimensional limit of
the full GMN integral equations is difficult for reasons explained in
\cite{GMN2}, and it is not straightforward to recover the
Atiyah-Hitchin metric in this approach. On the other hand, the
leading semiclassical contribution to the GMN metric studied here can
easily be continued to 
three-dimensions after the Poisson resummation described above.
We show that it reproduces the one-monopole contribution to the
Atiyah-Hitchin metric including the correct numerical prefactor.
As explained in \cite{Dorey1997}, this coefficient together with
the constraints of supersymmetry and other global symmetries
uniquely determines the Atiyah-Hitchin metric.

\section{Moduli Space, Wall Crossing Formula and BPS Spectrum}
\paragraph{}
To begin, let us review the relevant results in \cite{GMN1} and set the
subsequent notations.
We consider the pure $\cN=2$ supersymmetric gauge theory in $d=4$ dimensions
with gauge group $G=SU(2)$
and dynamical scale $\Lambda$.
This theory has a Coulomb branch $\cB$ where the complex adjoint
scalar field in the vector multiplet, $\phi$ acquires a VEV and
the $SU(2)$ gauge group is broken down to $U(1)$.
The massless bosonic fields on the Coulomb branch consist of a $U(1)$ gauge
field
and a complex scalar $a$ whose VEV (also denoted as $a$) parametrises
$\cB$ as a complex manifold. It is also convenient to define a
gauge-invariant order parameter $u=\langle {\rm Tr}\, \phi^{2} \rangle$
which provides a globally-defined coordinate on $\cB$.
\paragraph{}
The spectrum of the theory on the Coulomb branch $\cB$ contains BPS states
$\gamma=(n_{e},n_{m})$
carrying electric and magnetic charges, $n_e$ and $n_m$, under the
unbroken $U(1)$.
Each BPS state carries central charge $Z_{\gamma}(u)$
\begin{equation}\label{DefZgamma}
Z_{\gamma}(u)=\gae a(u)+\gam a_D(u)\,.
\end{equation}
which lies on a lattice in the complex plane with periods $a$ and
$a_{D}$. The magnetic period is determined by the
prepotential $\cF(a)$ \cite{SW1} via
\begin{equation}\label{Defmcharge}
a_D=\frac{\partial \cF(a)}{\partial a}\,.
\end{equation}
The prepotential $\cF(a)$ also determines the low-energy effective
gauge coupling:
\begin{equation}\label{Deftau}
\tau_{\rm eff}(a)=\frac{4\pi i}{g^2_{\rm eff}(a)}+
\frac{\Theta_{\rm eff}(a)}{2\pi}=
\frac{\partial^2
\cF(a)}{\partial a^2}\,.
\end{equation}
\paragraph{}
The exact low-energy action and BPS spectrum for the $SU(2)$ theory
were determined in \cite{SW1} by demanding a consistent realisation of
electric-magnetic duality on the moduli space $\cB$. In the exact
formulae the central charges $a(u)$ and $a_{D}(u)$ are identified as periods
of a meromorphic differential on a certain elliptic curve. In this
paper we will mostly be interested in the weak coupling regime where
$|a|\gg|\Lambda|$. In this case the effective coupling constant can be
approximated by its one-loop value,
\begin{eqnarray}
\tau_{\rm eff}(a) & \simeq & \frac{2i}{\pi} \log \frac{a}{\Lambda}\,,
\label{1loop}
\end{eqnarray}
up to corrections in powers of $(\Lambda/a)^{4}$ coming from
four-dimensional Yang-Mills instantons. To similar accuracy the
magnetic central charge is given as $a_{D}\simeq \tau_{\rm eff}a$.
\paragraph{}
The exact mass formula for BPS states of charge $\gamma$ is
$M_{\gamma}=|Z_{\gamma}|$ where the central charge $Z_{\gamma}$ defined
in (\ref{DefZgamma}). As mentioned in the introduction, the
Seiberg-Witten solution does not immediately specify the set of values of
$\gamma$ which are present in the theory.
Formally this corresponds to
determining the values of the second helicity supertrace
$\Omega(\gamma,u)=-\frac{1}{2}{\rm Tr}_{{\mathcal H}
{\rm BPS},\gamma}(-1)^{2J_3}(2J_3)^2$ at each point on the Coulomb branch.
Here $J_3$ is any generator of the rotation subgroup of the massive little
group, and this yields $\Omega(\gamma,u)=-2$ for the vector multiplet and
$\Omega(\gamma,u)=+1$ for half-hypermultiplet.
\paragraph{}
In the weakly coupled region $|a|\gg|\Lambda|$ the BPS spectrum can be
determined by semiclassical analysis.
It consists of the W-bosons of charges $\pm (1,0)$
and an infinite tower of Julia-Zee dyons $\pm(n,1)$ with unit magnetic charge
and arbitrary integer electric charge \footnote{In this paper we shall
follow the same normalization convention for
electric charges as in \cite{SW1}, which differs the convention used in
\cite{GMN1} by a factor of 2.} $n$. The plus and minus signs
correspond to charge conjugation. As we vary the modulus $u$, the
spectrum changes continuously except on the curves of real codimension one
where one or more BPS states becomes marginally stable.  The
degeneracies $\Omega(\gamma,u)$ can change discontinuously as we cross
such a ``wall of marginal stability''.
Following \cite{GMN1}, to
describe this phenomenon, we associate to each BPS state
of charge vector $\gamma$ a ray
$l_\gamma$ in
a complex $\zeta$-plane \footnote{The auxiliary
complex variable $\zeta$ is known as the spectral parameter. After adding
the point at infinity, $\zeta$ parametrises ${\mathbb{CP}}^1$.}:
\begin{equation}\label{Deflgamma}
l_\gamma :=\left\{\zeta: \frac{Z_\gamma(u)}{\zeta}\in \bR_{-}\right\}\,.
\end{equation}
These rays rotate in the $\zeta$-plane as we move in $\cB$.
At a point on the wall of marginal stability, a set of BPS rays
$\{Z_\gamma(u)\}$ become aligned,
and their charges can be parametrized as $\{N_1\gamma_1+N_2\gamma_2\},~N_1,N_2
> 0$,
for some primitive vectors $\gamma_{1,2}$ with $Z_{\gamma_1}/Z_{\gamma_2} \in
\bR_+$ \cite{GMN1}.
\paragraph{}
The ${\cal N}=2$ theory with gauge group $SU(2)$ offers a concrete
example of the wall-crossing phenomenon \cite{SW1}. The Coulomb branch $\cB$
corresponds to the complex plane with two singular points at
$u=\pm \Lambda^{2}$. The weak-coupling region
where $|u|\gg|\Lambda|^2$ is disconnected from the region near the origin
by a closed curve of marginal stability which passes through the
singular points.
As we enter the strongly coupled region near the origin in $\cB$, most of the
BPS states present in the semiclassical spectrum decay,
and we are left with only two BPS states:
They are the magnetic monopoles $\pm (0,1)$ and the dyons of unit electric
charges denoted either as $\pm (1,-1)$ or $\pm(1,1)$,
depending on whether we approach the curve of marginal stability from upper or
lower half plane in $\cB$.
Physically, taking the results in \cite{FB1} as an example, as we approach
curve of marginal stability from the upper half plane,
the W-bosons $\pm(1,0)$ decays into a (anti-) monopole $\pm(0,1)$ and a dyon of
charge $\pm (1,-1)$;
while a dyon $\pm (n,1)$ decays into $n+1$ (anti-) monopoles $\pm(0,1)$ and $n$
dyons of charge $\pm(1,-1)$.
Similar analysis can be done for the lower half plane.
\paragraph{}
Although the $SU(2)$ theory is well understood, theories with
additional matter and or larger gauge groups have a very complicated
set of walls of marginal stability.
Until recently,
a systematic determination of the BPS spectrum for generic $\cN=2$
SUSY gauge theories and all regions of the moduli space remained elusive.
In recent work Kontsevich and Soibelman \cite{KS1,KS2} have proposed a
wall-crossing formula which potentially solves this problem. Here we
will be mostly interested in the consequences of their conjecture for
the corresponding theory compactified to three dimensions. In the
following we briefly review the main results of \cite{GMN1}.
Focusing on four-dimensional $\cN=2$
Euclidean SYM with gauge group $SU(2)$,
we consider its compactification on the space of $\bR^3\times S^1$,
where $S^1 : x^4 \sim x^4+2\pi R$ has radius $R$. On length-scales much larger
than $R$, the
effective action on the Coulomb branch becomes three-dimensional. In
addition to the complex scalar $a$ of the four-dimensional theory,
there are now two additional real periodic scalar fields.
The first one is the electric Wilson line, which comes from the
component $A_{4}$ of the $U(1)$ gauge along $S^1$
denoted as
\begin{equation}\label{Defthetae}
\theta_e=\oint_{S^1} A_4 dx^4\,.
\end{equation}
Large gauge transformations shift the value of $\theta_{e}$ by integer
multiples of $2\pi$. Thus, we periodically identify
$\theta_{e}\sim \theta_{e}+2\pi$.
In three dimensions, we can also dualize the $U(1)$ Abelian gauge
field $A_i$, $i=1,2,3$ in favor of another real scalar
$\theta_{m}\in [0,2\pi]$ known as the magnetic Wilson line. This
appears in the classical action in the combination
$n_m\theta_{m}$ where $n_m$ is the total magnetic
charge. Dirac quantisation forces $n_m$ to take integer
values and, correspondingly, the theory is invariant under
shifts of $\theta_{m}$ by integer multiples of $2\pi$.
\paragraph{}
Taking into account the scalars $\{a,\bar{a},\theta_{e},\theta_{m}\}$,
the Coulomb branch of the compactified theory is a manifold ${\cal M}$
with real dimension four.
The low-energy effective field theory on the Coulomb branch is
then given by a $d=3$ sigma model with target $\cM$. The condition for
the effective theory to admit eight unbroken supercharges is for the
moduli space $\cM$ to be hyper-K\"ahler.
A hyper-K\"ahler manifold is K\"{a}hler with respect to
a triplet of complex strcutures $J_\alpha, \alpha=1,2,3$,
satisfying the $\mathfrak{su}(2)$ algebra $J_\alpha
J_\beta=-\delta_{\alpha\beta}+\epsilon_{\alpha\beta\gamma}J_\gamma$ and
$J_\alpha^2=-1$.
For each $J_\alpha$, there is a corresponding symplectic form $\omega_\alpha$,
and it is related to the metric $g$ on $\cM$ by $\omega_{\alpha}=-g
J_{\alpha}$.
More generally we can form linear combinations $c^{\alpha} J_\alpha$, with
$\sum_{\alpha=1}^3 c_\alpha^2=1$,
and the corresponding K\"ahler form is $c^\alpha \omega_\alpha$. The
manifold is therefore K\"ahler with respect to
a ${\mathbb{CP}}^1$ worth of complex structures.
Let the complex variable $\zeta$ parametrise such ${\mathbb{CP}}^1$,
the twistor space $\cT$ of a hyper-K\"ahler manifold is constructed so as to
incorporate all possible complex structures,
topologically $\cT \sim \cM \times {\mathbb{CP}}^1$ \cite{HKLR} \footnote{See
also \cite{IR} for a nice review on the twistor theory for hyper-K\"ahler
manifolds.}. In particular, choosing the complex coordinates holomorphic with
respect to $J_3$,
we can now organise the general K\"ahler form as
\begin{equation}\label{Defwzeta}
\w(\zeta)=-\frac{i}{2\z}\w_+ +\w_3-\frac{i}{2}\z \w_-\,,
\end{equation}
where we have defined $\w_{\pm}=\w_1\pm i\w_2$ and the metric $g$ is extracted
from $\zeta$ independent component of $\omega(\zeta)$.
The twistor space $\cT$ plays an important role in the construction of metric
on $\cM$ in \cite{GMN1}.
\paragraph{}
In the limit of large $R$, the low-energy effective action can be
obtained by dimension reduction of the four-dimensional low-energy
theory. To describe the action we define the complex combination
$z=\th_m-\tau_{\rm eff}\th_e$ which parametrises a torus
with complex structure $\tau_{\rm eff}(a)$.
In this limit the moduli space $\cal{M}$ corresponds to a fibration of
this torus over the Coulomb branch $\cB$ of the four-dimensional
theory. The real bosonic part of the
resulting
action is given in terms of scalar fields $\{a,\ba,\th_e,\th_m\}$ as
\begin{eqnarray}
\label{bosonic 3D action} \label{DefL3sf}
S_{\rm B}&=&\frac{1}{4}\int d^3 x \left(\frac{4\pi R}{g_{\rm eff}^2}
\partial_\mu a\,\partial^\mu\bar{a} +  \frac{g^2_{\rm eff}}{16\pi^3
R}\partial_\mu z\,\partial^\mu \bar
z\right)\,. \end{eqnarray}
In addition, surface terms give rise to pure imaginary
terms in the action depending on the total electric and magnetic
charge,
\begin{eqnarray}
S_{\rm Im} & = & i\left(n_{e}\,\,+\,\, \frac{\Theta_{\rm eff}}
{2\pi}\,n_{m}\right ) \,\theta_{e}\,\,+\,\,in_{m}\theta_{m}\,.
\label{Im}
\end{eqnarray}
The term proportional to $\Theta_{\rm eff}$ arises from dimensional
reduction of the $F\wedge F$ term in the low-energy action of the
four-dimensional theory after replacing $A_{4}$ by $\theta_{e}/2\pi
R$. The corresponding fermionic terms in the action take the form
\begin{eqnarray}
\label{fermionic 3D action}
S_{\rm F}&=&\frac{2\pi R}{g^2_{\rm eff}}\int d^3
x\left(i\bar\psi\bar\sigma^\mu\partial_\mu\psi+i\bar
\lambda\bar\sigma^\mu\partial_\mu \lambda\right)\,,
\end{eqnarray}
where $\lambda$ and $\psi$ are the dimensional reduction along the
$x_{4}$ direction of the four-dimensional Weyl fermions in the $U(1)$
vector multiplet whose lowest component is the scalar $a$.
\paragraph{}
Comparing with a general three-dimensional $\sigma$-model we can also
extract the leading $R\to \infty$ behavior of the hyper-K\"ahler metric
on $\cM$,
\begin{equation}\label{Defgsf}
g^{\rm sf}=R({\rm Im\, \tau_{\rm eff}})|da|^2+\frac{1}{4\pi^2 R}({\rm
Im}\, \tau_{\rm eff})^{-1}|dz|^2\,.
\end{equation}
This was called the ``semi-flat'' metric in \cite{GMN1}, as the two-torus
spanned by $\{\th_e,\th_m\}$ is flat.
The metric (\ref{Defgsf}) also makes apparent that $g^{\rm sf}$ is K\"ahler
with respect to the complex structure where $\{a,z\}$ are holomorphic
coordinates. Going to finite radius $R$, this semi-flat metric gets
corrected quantum mechanically by instanton contributions which
arise from the four-dimensional BPS states whose worldlines now wrap
around $S^1$. Until recently, no analytic formulae were available for
the exact metric except in the three-dimensional limit $R\rightarrow
0$ where it becomes the Atiyah-Hitchin metric \cite{SW3D}. The approach of
\cite{GMN1} seeks to determine the exact metric by insisting that the
total contribution of four-dimensional BPS states remains smooth
despite the discontinuous changes of the spectrum dictated by the
Kontsevich-Soibelman conjecture. We now review their main results.
\paragraph{}
The metric is effectively determined once the one-parameter family of
K\"ahler forms $\w(\zeta)$ introduced above is known. In particular
the K\"ahler form $\omega$ and metric $g$ are both derived from a
 K\"ahler potential $K$, with
$\omega = i\,\partial^2 K/(\partial z^{a} \partial z^{\bar b})d z^{a}\wedge d
z^{\bar b}$ and $g=2\,\partial^2 K/(\partial z^{a} \partial z^{\bar b})d z^{a}
d
z^{\bar b}=2g_{a\bar b}d z^{a} d z^{\bar b}$, respectively.
For any complex symplectic
manifold one can always find Darboux coordinates in which the
symplectic form becomes canonical. In the present case we introduce
complex coordinates $\mathcal{X}_{e}(\zeta)$ and $\mathcal{X}_{m}(\zeta)$,
in terms of which
\begin{equation}\label{Defwzeta2}
\omega(\zeta)=-\frac{1}{4\pi^2 R}\frac{d\mathcal{X}_{e}}{\mathcal{X}_{e}}
\wedge
\frac{d\mathcal{X}_{m}}{\mathcal{X}_{m}}\,.
\end{equation}
More generally, we also introduce a corresponding Darboux coordinate
$\cX_\gamma(\z)$ associated with any vector $\gamma$ in the
charge lattice determined by the relation
$\mathcal{X}_{\gamma_1+\gamma_2}=
\mathcal{X}_{\gamma_1}\mathcal{X}_{\gamma_2}$ where
$\mathcal{X}_{\gamma}= \mathcal{X}_{e}$ for $\gamma=(1,0)$ and
$\mathcal{X}_{\gamma}= \mathcal{X}_{m}$ for $\gamma=(0,1)$.
\paragraph{}
In the large-$R$ limit, the semi-flat metric (\ref{Defgsf})
corresponds to the choice,
\begin{equation}\label{defXsf}
\cX_{\gamma}^{\sf}(\z)=\exp\left(\pi R \z^{-1} Z_\ga+i\th_\ga+\pi R\bZ_\ga \z
\right)\,.
\end{equation}
It turns out that this asymptotic behavior, along with the discontinuities of
$\mathcal{X}_{\gamma}(\z)$ across the wall of marginal stability, as governed
by Konsevich-Soilbelman algebra
is enough to determine $\mathcal{X}_{\gamma}(\z)$
(and hence, the metric on $\cM$) at any point on the complex $\zeta$-plane.
Explicitly, the coordinate $\cX_\gamma(a,\th,\z)$ derived in \cite{GMN1} obeys
the following integral equation
\footnote{The expression inside the exponent differs from the GMN convention
\cite{GMN1} by a factor of $2$ as we divided electric charges by 2.}:
\begin{equation}\label{defXgamma}
\cX_{\ga}(\z)=
\cX^{\sf}_\ga(\z)\exp\left[-\frac{1}{2\pi i}
\sum_{\ga'\in\Gamma}\Omega(\ga';u)\langle\ga,\ga'\rangle\int_{l_{\ga'}}\frac{d\z'}{\z'}\frac{\z'+\z}{\z'-\z}\log\left(1-\sigma(\ga')\cX_{\ga'}(\z')\right)\right]\,,
\end{equation}
Let us define various quantities appeared above:
$\langle\ga,\ga'\rangle$ is the symplectic product between two charge
vectors, $\gamma=(\gae,\gam)$ and $\gamma'=(\gae',\gam')$,
which we can take to be
\begin{equation}
\langle\ga,\ga'\rangle =
\langle(\gae,\gam),(\gae',\gam') \rangle =-\gae\gam'+\gae'\gam\,,
\end{equation}
and $\sigma(\ga')=(-1)^{2\gae'\gam'}$, known as a ``quadratic refinement''.
The summation in (\ref{defXgamma}) is over the set of charges
$\Gamma$ in the theory and the BPS ray $l_{\ga'}$ associated with
$\gamma'$ is defined analogously
to (\ref{Deflgamma}).

\section{Semiclassical Limit of the Wall-Crossing Formula}
\paragraph{}
The integral equation (\ref{defXgamma}) for the
Darboux coordinate $\cX_{\ga}(\z)$ is of a standard type similar to
those arising in the context of the Thermodynamic Bethe
Ansatz \cite{Zam1990} \footnote{For recent interesting work relating the TBA
equation and wall-crossing formula, see \cite{Alexandrov}.}. Taking the
logarithm of this equation we see that the right hand
side contains a source term corresponding to the semi-flat expression
and an integral convolution. As usual we can solve the equation iteratively,
order by order in appropriate expansion parameter,
when the second term is smaller than the first. In \cite{GMN1} an
expansion of this sort was obtained at large radius $R|a|\gg1$, for any
point on the Coulomb branch. The
contributions of a BPS state with charge $\gamma$ is exponentially supressed by
a factor of $\exp(-2\pi R |Z_{\gamma}|)$ and, when $R|a|\gg1$, this factor
is small for all $\gamma$. Here we are instead interested in a weak
coupling expansion of the integral equation. Thus, we will restrict our
attention to the semiclassical region of the moduli space,
$|a|\gg\Lambda$, where $g^{2}_{\rm eff}\ll1$ while holding the
dimensionless quantity $R|a|$ fixed. As we explain below, the quantity
$\exp(-2\pi R |Z_{\gamma}|)$ is then supressed for all states with
non-zero magnetic charge.
In the weak coupling spectrum described above, this is the case for
all states except the massive gauge bosons $W^{\pm}$. This means we
must effectively resum the contributions from these states.
This is the main
difference between the weak-coupling expansion described below and the
large-$R$ expansion of \cite{GMN1}.
\paragraph{}
We begin by decomposing the Darboux coordinate $\cX_\ga(\z)$ as
\begin{equation}\label{DecomX}
\cX_\ga(\z)=\left[\cX_e(\z)\right]^{\gae}\left[\cX_m(\z)\right]^{\gam}\,,~~~\gamma=(\gae,\gam)\,.
\end{equation}
The integral equation (\ref{defXgamma})
for the electric and the magnetic Darboux coordinates is then given
as
\begin{eqnarray}
\cX_e(\z)&=&\cX_e^{\sf}(\z)\exp\left[-\frac{1}{2\pi
i}\sum_{\ga'\in\Gamma}c_e(\ga')\,\cI_{\ga'}(\z)\right]\,,
~~~c_e(\ga')=\Omega(\ga';u)\langle(1,0),\ga'\rangle\,,\label{DefXe}\\
\cX_m(\z)&=&\cX_m^{\sf}(\z)\exp\left[-\frac{1}{2\pi
i}\sum_{\ga'\in\Gamma}c_m(\ga')\,\cI_{\ga'}(\z)\right]\,,
~~~c_m(\ga')=\Omega(\ga';u) \langle(0,1),\ga'\rangle\,,\label{DefXm}
\end{eqnarray}
where $\cX_{e}^{\sf}(\z)$ and $\cX_{m}^{\sf}(\z)$ are given by (\ref{defXsf})
with $Z_\ga$ replaced by $Z_{e}$ and $Z_{m}$, respectively, and
$\cI_{\ga'}(\z)$ is defined to be
\begin{equation}\label{DefIzeta}
\cI_{\gamma'}(\z)=\int_{l_{\ga'}}\frac{d\z'}{\z'}\frac{\z'+\z}{\z'-\z}
\log(1-\sigma(\ga')\cX_{\ga'}(\z'))\,.
\end{equation}
Now, taking the weak coupling limit, to one loop order we find
\begin{equation}\label{appXsf}
\log\cX_e^{\sf}(\z)=\pi R a\z^{-1}+i\theta_e+\pi R \ba \z\,,
~~~\log\cX_m^{\sf}(\z)=\pi R a\tau_{\rm eff}(a)\z^{-1}+i\theta_m+\pi R
\overline{a\tau_\eff(a)} \z.
\end{equation}
We can see that in this limit $\log|\cX^{\sf}_m|\gg \log|\cX_e^{\sf}|$, this
has interesting consquences for deriving an iterative solution to
$\cX_\ga(\z)$.
Explicitly, let us expand $\log\cX_e(\z)$ and $\log \cX_m(\z)$ for the weak
coupling spectrum of $\cN=2$ $SU(2)$ gauge theory using (\ref{DefXe},
\ref{DefXm}):
\begin{eqnarray}
\log \cX_e(\z)&=&\log\cX_e^{\sf}(\z)-\frac{1}{2\pi i}\sum_{\gae'\in
\bbZ}\sum_{\gam'=\pm 1}c_e(\ga')\cI_{(\gae',\gam')}(\z)\,,\label{logXe}\\
\log \cX_m(\z)&=&\log\cX_m^{\sf}(\z)-\frac{c_m(W^{+})}{2\pi
i}\cI_{(1,0)}(\z)-\frac{c_m(W^{-})}{2\pi i}\cI_{(-1,0)}(\z)
\nonumber \\&-&\sum_{\gae'\in \bbZ}\sum_{\gam'=\pm 1}\frac{c_m(\ga')}{2\pi
i}\cI_{(\gae',\gam')}(\z)\,.
\label{logXm}
\end{eqnarray}
The central charge has the form $Z_\ga(a)= a(\gae+\gam \tau_{\rm eff}(a))$. The
mass of a
BPS particle $\gamma=(\gae,\gam)$ at weak-coupling limit is given by
\begin{equation}
| Z_{\ga}(a)|= |a|\sqrt{\left(\gam
\frac{4\pi}{g^2_{\rm eff}}\right)^2+\left(\gae+\gam\frac{\Theta_{\rm
eff}}{2\pi}\right)^2},
\end{equation}
where $g_\eff$ and $\Theta_\eff$ now denote the effective coupling constant and
the
effective vacuum angle. The BPS spectrum $\Gamma$ consists of the $W$-bosons of
charges $\pm(1,0)$
which we denoted as $W^{\pm}$ and only contribute to the middle two terms in
(\ref{logXm}),
and the remaining summations in (\ref{logXe}) and (\ref{logXm}) are over the
infinite tower of dyons with charges $\pm (n,1)$, $n\in {\mathbb Z}$.

\paragraph{}
Let us further describe our weak coupling, iterative approach to solving $\log
\cX_\ga$.
At the leading order, we substitute the semi-flat coordinates (\ref{appXsf})
into the right hand side of (\ref{logXe}, \ref{logXm}) and ignore the
components which vanish as $g_\eff\rightarrow 0$. As the result,
the magnetic coordinate $\cX_m$ receives additional order one contribution due
to the $W$-bosons, while the dyon contributions to $(\cX_e, \cX_m)$ are
exponentially suppressed as $\sim \exp(-c/g_\eff^2)$ along the integration
contours
$\{l_\ga\}$.
We shall denote the resultant coordinates at this order as
$(\cX^{(0)}_e,\cX^{(0)}_m)$. Thus we have
\begin{eqnarray}
\log\cX_e^{(0)}(\z)&=&
\log\cX_e^{\sf}(\z)\,,~~~\log\cX_m^{(0)}(\z)=\log\cX_m^{\sf}(\z)+\log\cD(\z)\,,\label{DefXe0Xm0}\\
\log\cD(\z)&=&-\frac{c_m(W_+)}{2\pi
i}\int_{l_{W^+}}\frac{d\z'}{\z'}\frac{\z'+\z}{\z'-\z}
\log\left(1-\cX_e^{\sf}(\z')\right)\nonumber\\
&-&\frac{c_m(W_{-})}{2\pi
i}\int_{l_{W^-}}\frac{d\z'}{\z'}\frac{\z'+\z}{\z'-\z}
\log\left(1-1/\cX_e^{\sf}(\z')\right)\,,\label{DefDz}
\end{eqnarray}
where the BPS rays for the W-bosons are defined as $l_{W^{\pm}}:=\{\z: \pm a/\z
\in {\mathbb R}_{-}\}$.
We can now expand $(\cX_e(\z),\cX_m(\z))$ into
\begin{equation}
\log\cX_{e}(\z)\,\,\,=\,\,\,\log\cX_{e}^{(0)}(\z)\,\,+\,\,\delta
\log\cX_e(\z)\,,
\quad{}
\log\cX_{m}(\z)\,\,\,=\,\,\,\log\cX_{m}^{(0)}(\z)\,\,+\,\,\delta
\log\cX_m(\z)\,.
\end{equation}
By plugging $(\cX_e^{(0)}(\z),\cX_m^{(0)}(\z))$ into (\ref{logXe}) and
(\ref{logXm}),
we can read off the sub\-leading-order corrections to the coordinates:
\begin{eqnarray}
\delta\log\cX_e(\z)&=&-\frac{1}{2\pi i}\sum_{\gae'\in \bbZ}\sum_{\gam'=\pm 1}
c_e(\ga')\cI_{(\gae',\gam')}^{(0)}(\z)\,,\label{delogXe0}\\
\delta \log\cX_m(\z)&=&-\frac{1}{2\pi i}\sum_{\gae'\in
  \bbZ}\sum_{\gam'=\pm 1} c_m(\ga')\cI_{(\gae',\gam')}^{(0)}(\z)\,,
\label{delogXm0}
\end{eqnarray}
where we have defined the short-hand notation for the integral:
\begin{equation}
\cI_{(\gae',\gam')}^{(0)}(\z)
=\int_{l_{\ga'}}\frac{d\z'}{\z'}\frac{\z'+\z}{\z'-\z}
\log\left(1-\left[\cX_e^{(0)}(\z')\right]^{\gae'}\left[\cX_m^{(0)}(\z')\right]^{\gam'}\right)\,.\label{DefI0}
\end{equation}
We will soon see that these terms generate the exponentially
suppressed dyon contributions.
\paragraph{}
To extract the corresponding correction to the metric on $\cM$,
we compute the symplectic form $\w(\z)$ (\ref{Defwzeta2}),
including the corrections (\ref{delogXe0}) and (\ref{delogXm0}) to
(\ref{DefXe0Xm0}).
The result yields
\begin{eqnarray}\label{omegaexpansion}
\w(\z)&\approx&-\frac{1}{4\pi^2 R} d(\log\cX_e^{(0)}(\z)+\delta\log \cX_e (\z))
\wedge d(\log\cX_m^{(0)}(\z)+\delta\log \cX_m (\z))\nonumber\\
&=& \w^{\sf}(\z)+\w^{W}(\z)+\w^{\rm dyon}(\z)+{\mathcal{O}}(\delta^2)\,,
\end{eqnarray}
where in the first line we have re-written $\cX_\ga(\z)$ in terms of
$\cX_e(\z)$ and $\cX_m(\z)$ using (\ref{defXgamma}).
The various terms in (\ref{omegaexpansion}) are then given by
\begin{eqnarray}
\w^{\sf}(\z)&=&-\frac{1}{4\pi^2 R}d\log\cX_e^{\sf}(\z)\wedge d\log
\cX_m^{\sf}(\z)\,,\label{wsf}\\
\w^{W}(\z)&=&-\frac{1}{4\pi^2 R}d\log\cX_e^{\sf}(\z)\wedge d\log
\cD(\z)\,,\label{wW}\\
\w^{\rm dyon}(\z)&=&-\frac{1}{4\pi^2 R}\left(d\delta\log\cX_e(\z)\wedge
d\log\cX_m^{(0)}(\z)+d\log\cX_e^{(0)}(\z)\wedge
d\delta\log\cX_m(\z)\right)\,.\label{winst}
\end{eqnarray}
We will evaluate the integral expressions $\w^{W}(\z)$ and $\w^{\rm dyon}(\z)$
in turns. The term $\w^{W}(\z)$ corresponds to the one-loop perturbative
correction to the metric due the $W^{\pm}$ boson.
Its evaluation is essentially the same as the computation done
in the section 4.3 of \cite{GMN1} where the contribution of a single
electrically charged state was considered
(see equation (4.39) there). Using these results, we obtain:
\begin{eqnarray}
\w^{W}(\z)&=&-\frac{1}{4\pi^2 R}d\log\cX_e^{\sf}(\z)\wedge \left[2\pi i
A^{W}(a,\ba)+\pi i V^{W}(a,\ba)(\z^{-1} da-\z d\ba)\right]\label{expwW}\,,\\
A^W(a,\ba)&=&\frac{R}{\pi}\sum_{k\neq 0}|a| e^{ik\th_e} K_1(2\pi R|ka|)
\left(\frac{da}{a}-\frac{d\ba}{\ba}\right)\,,\label{AW}\\
V^{W}(a,\ba)&=&-\frac{2R}{\pi}\sum_{k\neq 0}  e^{in\th_e}K_0(2\pi
R|ka|)\,,\label{VW}
\end{eqnarray}
where $K_{\nu}(x)$ are modified Bessel functions of the second kind. In the
computation, we have also used the facts $\Omega(W^{\pm})=-2$ as W-bosons
belong to vector multiplet and
$\langle(0,1),(\pm 1,0)\rangle=\pm 1$ to set $c_m(W_\pm)=\mp 2$.
In \cite{GMN1}, the authors further considered the large radius limit $R\gg
1/|\Lambda|$ and further allowed $2\pi R |a| \gg 1$,
this sets $K_{\nu}(2\pi R |ka|) \sim e^{-2\pi R |ka|}$,
and the series in (\ref{AW}) and (\ref{VW}) can be naturally interpreted as
summing up exponentially suppressed, instanton-like, contributions.
As mentioned above, we are taking weak coupling limit $|a/\Lambda| \gg 1$ while
keeping $2\pi R |a|$ fixed and arbitrary. To go to small values of
$R|a|$, one instead needs to Poisson resum the series of Bessel functions
\cite{Ooguri Vafa}. The resulting geometry corresponds to a finite
shift of the coupling constant \cite{SeiShe}:
\begin{equation}\label{DefMn}
\frac{2\pi R}{g^2_{\rm eff}}\rightarrow\frac{2\pi R}{g_{\rm eff}^2}-
\sum_{n\in \bbZ}\frac{1}{2\pi
| M(n)|}\,,
~~~|M(n)|=\sqrt{|a|^2+\left(\frac{\theta_e}{2\pi R}+\frac{n}{R}\right)^2}\,,
\end{equation}
which determines the harmonic function $V$ in the Gibbons-Hawking
ansatz for the hyper-K\"ahler metric (see eq.\ (4.2) of \cite{GMN1}).
In the limit $R\to0$, this reduces to the shift $1/e_{\rm eff}^2\rightarrow
1/e_{\rm eff}^2-1/(2\pi M_W)$ where $1/e_{\rm eff}^2=2\pi R/g_{\rm eff}^2$ and
$M(0)$ are the gauge coupling
and the mass of $W$-boson in three dimensions \cite{SW3D, Dorey1997}.

\paragraph{}
Now for the more complicated $\w^{\rm dyon}(\z)$, our strategy here is to
evaluate the integrals involved by saddle point approximation,
as they are dominated by the exponential terms at weak coupling.
Explicitly we have the following series for $\w^{\rm dyon}(\z)$ in
(\ref{winst}):
\begin{eqnarray}
\w^{\rm
dyon}(\z)&=&\sum_{\gamma'=(n_e',\pm 1)}\w_{\ga'}(\z)\,,\\
\w_{\ga'}(\z)&=&-\frac{1}{4\pi^2
R}\frac{d\cX_{\ga'}^{(0)}(\z)}{\cX_{\ga'}^{(0)}(\z)}\wedge\left(\frac{1}{2\pi i
}\int_{l_{\ga'}}\frac{d\z'}{\z'}\frac{\z'+\z}{\z'-\z}
\frac{\cX^{(0)}_{\ga'}(\z')}{1-\cX^{(0)}_{\ga'}(\z')}\frac{d\cX_{\ga'}^{(0)}(\z')}{\cX_{\ga'}^{(0)}(\z')}\right)
\nonumber\\
&\approx & -\frac{1}{4\pi^2
R}\frac{d\cX_{\ga'}^{\sf}(\z)}{\cX_{\ga'}^{\sf}(\z)}\wedge\left(\frac{1}{2\pi i
}\int_{l_{\ga'}}\frac{d\z'}{\z'}\frac{\z'+\z}{\z'-\z}\frac{\cX^{(0)}_{\ga'}(\z')}{1-\cX^{(0)}_{\ga'}(\z')}
\frac{d\cX_{\ga'}^{\sf}(\z')}{\cX_{\ga'}^{\rm
sf}(\z')}\right)\,,
\label{Defwemk}
\end{eqnarray}
where we have used the fact that $\Omega(\ga',a)=1$ for all the dyon states.
Along each integration contour
$l_{\gamma'} : Z_{\gamma'}/\z' \in {\mathbb{R}_{-}}$, the zeroth order Darboux
coordinate
$\cX^{(0)}_{\ga'}(\z')=\cX^{\sf}_{\ga'}(\z')\cD(\z')$ is proportional to
exponential factor
$\exp\left[-\pi R|Z_{\ga'}|(|\z'|+1/|\z'|)\right]$, which ensures the
convergence of the integral.
At the weak coupling $|\tau_\eff| \gg 1$ and $|Z_{\ga'}|\gg 1$,
we can therefore Taylor-expand
$\cX_{\ga'}^{(0)}(\z')/(1-\cX_{\ga'}^{(0)}(\z'))$ in the integrand above into
\begin{equation}\label{expandX}
\frac{\cX_{\ga'}^{(0)}(\z')}{1-\cX_{\ga'}^{(0)}(\z')}=\sum_{k=1}^{\infty}\left[\cX_{\ga'}^{(0)}(\z')\right]^k\,,~~~\z'
\in l_{\ga'}\,,
\end{equation}
and perform saddle point analysis for each term in the expansion.
Here we have also further approximated $d\cX^{(0)}(\z)/\cX^{(0)}(\z)$ by
$d\cX_{\ga}^{\sf}(\z)/\cX_{\ga}^{\sf}(\z)$,
as the contribution proportional to $d\cD(\z)/\cD(\z)$ is of higher order in
$g^2_{\rm eff}$ in our saddle point analysis.
The saddle point analysis amounts to extremizing $\exp[-k\pi R |Z_{\ga'}|(
| \z'|+1/|\z'|)]$ with respect to $|\z'|$ in each term of (\ref{expandX}), we
can then deduce that the saddle point sits at
$\z'=-Z_{\gamma'}/|Z_{\gamma'}|$.
Upon substitution and performing the Gaussian fluctuation integral,
the leading expression for $\w_{(\ga',k)}(\z)$ is given by
\begin{eqnarray}
\w^{\rm dyon}(\z)&=&\sum_{\gamma'=(n_e',\pm
1)}\sum_{k=1}^{\infty}\w_{(\ga',k)}(\z)\,,\label{wdyon1}\\
\w_{(\ga',k)}(\z)&=&
\cJ_{(\ga',k)}\frac{d\cX_{\ga'}^{\sf}(\z)}{\cX_{\ga'}^{\sf}(\z)}\wedge
\left[{|Z_{\ga'}|}\left(\frac{dZ_{\ga'}}{Z_{\ga'}}-\frac{d\bar{Z}_{\ga'}}{\bar{Z}_{\ga'}}\right)
-\left(\frac{dZ_{\ga'}}{\z}-\z d\bar{Z}_{\ga'}\right)\right]\, ,\\
\cJ_{(\ga',k)}&=& -\frac{1}{8\pi^2 i }
\frac{[\cD(- e^{i\phi_{\ga'}})^{\gam'}]^k}{\sqrt{k R |Z_{\ga'}|}}
\exp\left[k(-2\pi R|Z_{\ga'}|+i\th_{\ga'})\right]\,,\label{sadcJgmk}
\end{eqnarray}
\begin{equation}
\begin{aligned}
\log\cD(-e^{i\phi_{\ga'}})^{\gam'}
\approx \pm\log \cD(\mp i)=
\frac{1}{\pi i}\int^{\infty}_{0} \frac{dy}{y}\left[\frac{y+ i}{y-
i}\log\left(1-e^{-\pi R|a|(y+1/y)+i\theta_e}\right) \right. \\ \left.
-\frac{y- i}{y+ i}\log\left(1-e^{-\pi R|a|(y+1/y)-i\theta_e}\right)\right]\,.
\label{sadcD}
\end{aligned}
\end{equation}
Here $e^{i\phi_{\ga'}}=\frac{\gae'+\tau_\eff\gam'}{|\gae'+\tau_\eff\gam'|}$,
and in the weak coupling limit we have further approximated $e^{i\phi_{\ga'}}
\approx i\gam'$, this is consistent with the saddle point
approximation and
ensures the resultant one loop determiant $\cD$ being real.
By substituting $y=e^t$ and using $\gam'=\pm 1$, the equation (\ref{sadcD}) can
be
re-expressed as
\begin{eqnarray}
\pm \log\cD(\mp i)
=\frac{2}{\pi }\int^{\infty}_{0} \frac{dt}{\cosh t}\left[\log\left(1-e^{-2\pi
R|a|\cosh t+i\theta_e}\right)
+\log\left(1-e^{-2\pi R|a|\cosh t-i\theta_e}\right)\right]\,.\label{sadcD2}
\end{eqnarray}
In the next section we will show that this expression precisely
corresponds the ratio of one-loop determinants corresponding to
small fluctuations around a classical dyon contribution.
\paragraph{}
Now we can extract the correction to the symplectic form on the moduli space
$\cM$ from the $\z$-independent part of $\w^{\rm dyon}(\z)$. As $g_\eff^2 \to
0$, keeping only the leading terms with $k=1$ and focusing on only instanton
(i.e. $\gam'= +1$)
contributions, we obtain
\begin{eqnarray}\label{explicitw3inst}
\w^{\rm inst.}_{3}
&=&\sum_{\ga'=(\gae',1)}
\cJ_{(\ga',1)}\left((2\pi R) dZ_{\ga'}\wedge d\bar{Z}_{\ga'}
+i|Z_{\ga'}|d\th_{\ga'}\wedge\left(\frac{dZ_{\ga'}}{Z_{\ga'}}-\frac{d\bar{Z}_{\ga'}}{\bar{Z}_{\ga'}}\right)\right)\nonumber\\
&=&\sum_{\ga'=(\gae',1)}
\cJ_{(\ga',1)}\left(2\pi R|\gae'+\tau_\eff(a)|^2 da\wedge d\bar{a}
+i|\gae'+\tau_\eff(a)|d\th_{\ga'}\wedge |a|
\left(\frac{da}{a}-\frac{d\bar{a}}{\bar{a}}\right)\right)\,,\nonumber\\
\end{eqnarray}
where $\w^{\rm dyon}_{3}=\w^{\rm inst.}_{3}+\bar \w^{\rm inst.}_{3}$ ($\bar
\w^{\rm inst.}_{3}$ corresponds to $\gam'=-1$ contributions). Here in the
second line of (\ref{explicitw3inst}) we have used the weakly coupled
expression
for the central charge $Z_{\ga'}= a(\gae'+\tau_\eff(a))$.
Explicitly, let us write out the $g_{a\ba}$ component from
(\ref{explicitw3inst}),
which gives the dominant contribution in weak-coupling expansion:
\begin{equation}\label{gaainst1}
g_{a\bar{a}}^{\rm
inst.}=\frac{\sqrt{R}}{4\pi}\sum_{\ga'=(\gae',1)}\frac{\cD(-i)
| Z_{\ga'}|^{3/2}}{|a|^2}\exp\left(-2\pi R|Z_{\ga'}|+ i\th_{\ga'}\right)\,.
\end{equation}
Other metric components $g^{\rm inst.}_{a\bar{z}}, g^{\rm inst.}_{\ba z}$ which
are suppressed by $g^2_\eff$ can also be readily extracted from
(\ref{explicitw3inst}).
By including these additional metric components and using the complex
coordinates $(z,\bar{z})$ introduced
earlier we can calculate a K\"ahler potential, $K^{\rm dyon}$
corresponding to the K\"ahler form  $\w^{\rm dyon}_{3}$,
\begin{equation}
\theta_m=\frac{1}{2}\left[ (z+\bar z)+\frac{i\,{\rm Re}\,\tau_\eff}{{\rm
Im}\,\tau_\eff}(z-\bar z) \right]\,, \quad \theta_e=\frac{i}{2\,{\rm
Im}\,\tau_\eff}(z-\bar z)\,,
\end{equation}
we recover the K\"ahler potential for the K\"{a}hler form
(\ref{explicitw3inst}):
\begin{equation} \label{Kcharges}
K^{\rm dyon}=\sum_{\ga'=(\gae',\pm 1)} \frac{\cD(-i)}{4\pi^3 R^{3/2}
\sqrt{|Z_{\gamma'}|}}\exp(-2\pi R|Z_{\gamma'}|+i\theta_{\ga'})\,.
\end{equation}
\paragraph{}
Finally we further expand the metric (\ref{gaainst1}) at weak coupling as
\begin{eqnarray}
g_{a\bar{a}}^{\rm inst.}&\approx
&\frac{\sqrt{R}}{4\pi}\left(\frac{4\pi}{g^2_\eff}\right)^{3/2}
\sum_{\ga'=(\gae',1)}\frac{\cD(-i)}{|a|^{1/2}}\exp\left( -S_{\rm
Mon}-S_{\varphi}^{(\gae')}\right)\,,\label{gaaApprox}\\
S_{\rm Mon.}&=&\frac{8\pi^2 R}{g^2_\eff}|a|-i \theta_m\,,\label{Smon}\\
S_{\varphi}^{(\gae')}&=&\frac{g^2_\eff R
| a|}{4}\left(\gae'+\frac{\Theta_\eff}{2\pi}\right)^2-i\gae'\theta_e\,.\label{Schi}
\end{eqnarray}
The exponent is the Euclidean action of a magnetic monopole
thought of as a static field configuration on
$\mathbb{R}^{3}\times S^{1}$. The remaining term
$S_{\varphi}^{(\gae')}$ is the leading contribution from the electric
charge of the dyon. In four dimensional theory, the corresponding
contribution to the dyon mass to comes from the
slow motion of the monopole in the $S^{1}$ factor of its moduli
space \cite{tw}. The further shift of the electric charge
$\gae'\to \gae'+\Theta_{\rm eff}/2\pi$ corresponds to the
familiar Witten effect \cite{wt}.
\paragraph{}
One additional subtlety described in Section 4.1 of \cite{GMN1}
concerns the appropriate choice of coordinates on the torus fibre of
the moduli space. When working near a singularity in the moduli space where
a ratio of BPS particle masses vanishes it is appropriate to change
variables to a coordinate which is single-valued in a neighbourhood of
the singular point. In the present case we are interested in the
semiclassical region of the moduli space near infinity and the
singularity corresponds to the logarithm in the one-loop effective coupling
(\ref{1loop}). Adapting eq.\ (4.13) of \cite{GMN1} to this case,
  the corresponding change of variable is
\begin{eqnarray}
\theta_{m} & \rightarrow &
\theta_{m}'\,\,=\,\,\theta_{m}\,+\,\frac{\Theta_{\rm
    eff}}{2\pi}\,\theta_{e}
\nonumber
\end{eqnarray}
Implementing this replacement, equation (\ref{Schi}) becomes
\begin{eqnarray}
S_{\varphi}^{(\gae')}&=&\frac{g^2_\eff R
| a|}{4}\left(\gae'+\frac{\Theta_\eff}{2\pi}\right)^2-i\left
(\gae'+\frac{\Theta_\eff}{2\pi}\right)\theta_e\,.
\label{Schi2}
\end{eqnarray}
and we see that the imaginary part of the action
$S_{\rm Mon.}+S_{\varphi}^{(\gae')}$ agrees with eqn. (\ref{Im}) obtained
directly from dimensional reduction of the four-dimensional metric.

\paragraph{}
The hyper-K\"ahler metric on $\cal{M}$ completely determines the terms
in the low-energy
effective action for the massless fields with at most two derivatives
or four fermions. These terms correspond to a three-dimensional
supersymmetric sigma model  with target space $\cM$;
\begin{equation}\label{3DSeff}
S_{\rm eff}^{\rm (3D)}=\frac{1}{4} \int d^3 x \left(g_{ij}(X)\left[\partial_\mu
X^i\partial^{\mu} \bar{X}^{j}+i\bar{\Omega}^{i}  \ssl{D}\Omega^j \right]
+\frac{1}{6} R_{ijkl}(\bar{\Omega}^{i}\cdot\Omega^k)(\bar{\Omega}^{j}\cdot
\Omega^l)\right)\,.
\end{equation}
Here $\{X^i\}$ are four bosonic scalar fields and
$\{\Omega_i^\alpha\}$ are their Majorana fermionic superpartners.
\paragraph{}
As usual, the bosonic scalar fields $\{X^{i}\}$ define the coordinates on the
quantum moduli space $\cM$. In the semiclassical limit, the metric
goes to its semiflat value and, choosing
complex coordinates $\{a,\bar{a},z,\bar{z}\}$, the bosonic action must
coincide with (\ref{DefL3sf}). If we choose appropriately rescaled coordinates
defined by
\begin{equation}\label{DefX1X2}
a=\frac{g_\eff}{2\sqrt{\pi R}} X^1 \,,~~~z=\frac{4\pi \sqrt{\pi
R}}{g_\eff}X^2\,,
\end{equation}
then the semi-flat metric $g^{\sf}$ (\ref{Defgsf}) simply reduces to
the flat metric
$\delta_{i\bar{j}}\,,~~i,{j}=1,2$ in this limit.
\paragraph{}
By comparing the fermionic terms in the action with
(\ref{fermionic 3D action}),
we can rewrite the action in terms of the
Weyl spinors $\{\lambda_\alpha, \bar{\lambda}_{\dot{\alpha}},
\psi_\alpha,\bar{\psi}_{\dot{\alpha}}\}$ of four-dimensional $U(1)$
vector multiplet. Following \cite{Dorey1997}, we rewrite the latter in
terms three-di\-men\-sional Majorana fermions
$\{\chi^{\bar{a}},\bar{\chi}^{a}\}$ via
\begin{equation}\label{4D3Dfermions}
\lambda_{\alpha}=\chi^{\bar{1}}_{\alpha}\,,~~\epsilon_{\alpha\dot{\beta}}\bar{\lambda}^{\dot{\beta}}=\bar{\chi}^1_{\alpha}\,,~~
\psi_{\alpha}=\chi^{\bar{2}}_{\alpha}\,,~~\epsilon_{\alpha\dot{\beta}}\bar{\psi}^{\dot{\beta}}=\bar{\chi}^2_{\alpha}\,.
\end{equation}
The Majorana fermions $\{\Omega^i\}$ appearing in (\ref{3DSeff}) can be
then be expressed as
\begin{equation}\label{Omegachi}
\Omega^i_{\alpha}=M^{i c}(X)
\bar{\chi}_{c\alpha}\,,~~~\Omega^{\bar{i}}_{\alpha}=M^{\bar{i} \bar{c}}(X)
\chi_{\bar{c}\alpha}\,,
~~~c=1, 2\,,
\end{equation}
where $M^{i c}(X)$ and $M^{\bar{i}\bar{c}}(X)$ are undetermined
matrices which can depend non-trivially on the bosonic scalars $X^i$.
Matching with the fermion kinetic terms in (\ref{fermionic 3D action})
imposes the normalisation condition,
\begin{equation}\label{normalization}
\delta_{i\bar{j}}M^{i a}(X) M^{\bar{j}\bar{b}}(X)=\left(\frac{8\pi
R}{g^2_\eff}\right)\delta^{a\bar{b}}\,.
\end{equation}
In a vacuum where $\theta_{e}=0$, the relation between the fermion
bilinears appearing in (\ref{3DSeff}) and the four dimensional
fermions can be made explicit as,
\begin{equation}\label{2fermions}
\bar{\Omega}^{1}\cdot\Omega^{1}=\left(\frac{8\pi
R}{g^2_\eff}\right){\lambda}\cdot\bar{\lambda}\,,~~~
\bar{\Omega}^{2}\cdot\Omega^{2}=\left(\frac{8\pi
R}{g^2_\eff}\right){\psi}\cdot\bar{\psi}\,.
\end{equation}
\paragraph{}
The four-fermion term in the action (\ref{3DSeff}) involves the
Riemann tensor of the hyper-K\"{a}hler metric on $\cal{M}$.
The leading semilcassical computation is computed using our result
(\ref{gaaApprox}) for the metric.
As the metric is K\"ahler in the complex coordinates $\{a, \ba, z, \bar{z}\}$,
we conclude that at the leading order in $g$ expansion, the only non-vanishing
components, up to
the symmetries of the Riemann tensor, are
\begin{equation}\label{RiemannTensor1}
R_{a\bar{z}z\ba}=R_{a\ba z\bar{z}}=g_{a\bar{p}}\partial_z (g^{\bar{p}
q}\partial_{\bar{z}} g_{\ba q})\,,~~p,q =\{a\,, z\}\,,
\end{equation}
and we can reduce the expressions above to
\begin{equation}\label{expRiemanntensor}
R_{a\bar{z}z\ba}=R_{a\ba z\bar{z}}=-\frac{1}{4} g_{a\ba}^{\rm inst.}\,.
\end{equation}
In terms of $\{X^{i},X^{\bar{j}}\}$ defined in (\ref{DefX1X2}),
we can relate the Riemann tensor (\ref{RiemannTensor1}) extracted from the
integral formula in \cite{GMN1} with the one in new coordinates via
\begin{equation}\label{Riemann2}
R_{1\bar{2}\bar{1}2}=\left|\frac{d a}{d X^1}\frac{d z}{d X^2}\right|^2
R_{a\bar{z}\bar{a}z}=(2\pi)^2 R_{a\bar{z}\bar{a}z}\,.
\end{equation}
The Riemann tensor captures the
quantum corrections to the metric, both perturbatively and non-perturbatively.
Now we can use the above conversion between $\Omega^{1,2}$ and
$\lambda,\psi$
(\ref{2fermions}) to
extract the prediction for the four fermion in the low-energy effective
Lagrangian from (\ref{gaaApprox}).
After taking into account the symmetries of the Riemann tensor,
and restricting again to the leading $k=\gam'=1$ sector,
we obtain
\footnote{Strictly speaking the four-fermion vertex is only correct as written
in a vacuum where $\theta_{e}=0$. For $\theta_{e}\neq 0$, the matrices
appearing in (\ref{Omegachi}) effect a rotation which changes the
chirality of the vertex but preserves the overall normalisation which
is subject to (\ref{normalization}). We will supress this subtlety in
the following.},
\begin{equation}\label{4fermi}
S_{\rm 4F}=\frac{2^{9/2}\pi}{R|a|^{1/2}} \left(\frac{2\pi
R}{g^2_{\rm eff}(a)}\right)^{7/2}\cD(-i)
\exp\left[-S_{\rm Mon}\right]
\sum_{\gae'\in\bbZ}\exp\left[-S_{\varphi}^{(\gae')}\right]\int d^3
x(\psi\cdot\bar\psi)(\lambda\cdot\bar\lambda)\,.
\end{equation}
We shall next verify this term in the effective action via a
direct semiclassical calculation.

\section{Semiclassical Instanton Calculation}
\paragraph{}
In this Section we will compute the monopole and dyon contributions to
the action from first principles. We focus on the leading contribution
of magnetic charge $\gam=k=1$ and arbitrary
electric charges $\gae \in \bbZ$ to an appropriate four fermion correlator to
be defined momentarily. A similar calculation in the three-dimensional
limit was performed in \cite{Dorey1997} and for the corresponding theory
with 16 supercharges on $\mathbb{R}^{3}\times S^{1}$ in
\cite{Dorey2000A, Dorey2000B}. We will refer the reader to these references
for some of the details.
\paragraph{}
We begin by considering
a static BPS monopole of the $\cN=2$ theory in four-dimensional
Minkowski spacetime. The bosonic moduli of this soliton consist of three
coordinates $X^{1,2,3}$ specifying the position of its centre in
$\mathbb{R}^{3}$ and a global $U(1)$ charge angle parametrized by
$\varphi\in [0,2\pi]$. The moduli space is thus ${\mathbb R}^3\times
S^1_{\varphi}$.
As the configuration is one-half BPS, there are
four fermionic zero modes in the monopole background, as
generated by half of the eight supercharges.
It is convenient to work in a formalism where monopoles are preserved by
supercharges of the same four-dimensional chirality,
to do so we embed the monopole as self-dual field configuration in
an auxiliary four-dimensional ${\cal N}=2$
gauge theory where the scalar field in
the BPS equation arises corresponds to a component of the
four-dimensional gauge field. In general, this is not the same as the
four-dimensional theory we compactified on $\mathbb{R}^{3}\times
S^{1}$ in the previous section where the corresponding field remains a
scalar. However, as we discuss below,
the fermions of the two theories are related to each
other by an $R$-symmetry rotation (see \cite{Dorey2000A}).
\paragraph{}
The left and right handed Weyl fermions of the auxiliary theory are
denoted $\rho^{A}_{\delta}$ and $\bar{\rho}^{A}_{\dot{\delta}}$
respectively where $A=1,2$ and $\delta,\,\dot{\delta}=1,2$.
In terms of these fermions the four zero modes of the instanton are
all left-handed yielding a
non-zero contribtion to the correlator,
\begin{equation}
\cG_{4}(\by_1,\by_2,\by_3,\by_4)=\langle{\prod_{A=1}^2
\rho^{ ~ A}_1(\by_{2A-1})\rho^{~
A}_2(\by_{2A})}\rangle\,.
\label{corr}
\end{equation}
corresponding to a vertex of the form
$(\bar{\rho}^{1}\cdot\bar{\rho}^{1})\,\,
(\bar{\rho}^{2}\cdot\bar{\rho}^{2})$ in the low-energy effective
action. The fermions of the auxiliary theory are related
to the original four-dimensional Weyl fermions by an $SO(3)$
$R$-symmetry rotation which mixes left and right-handed
chiralities but preserves the normalisation of the four-fermion
vertex in the effective Lagrangian.
In a vacuum where $\theta_{e}=0$ the zero modes of a monopole
are chirally symmetric in the original four-dimensional theory,
and the explict relation takes the form
\footnote{As in the previous section, when $\theta_{e}\neq 0$, the rotation
leads to chirally asymmetric
vertex when written in terms of $\lambda$ and $\psi$.},
\begin{eqnarray}
(\bar{\rho}^{1}\cdot\bar{\rho}^{1})(\bar{\rho}^{2}\cdot\bar{\rho}^{2}) & = &
(\psi\cdot\bar\psi)(\lambda\cdot\bar\lambda)\,.
\label{transfmn}
\end{eqnarray}

\paragraph{}
In the weak-coupling approximation we can
replace the fermions in the correlation function (\ref{corr})
with their zero mode
values multiplied by corresponding
Grassmann collective coordinates $\xi^{A}_{\delta}$. The explicit form
of the zero modes is given in Appendix C of \cite{Dorey1997}.
As we are interested in comparing with the low-energy
effective action we focus on the large-distance limit of the
correlation function and of the fermion zero modes. We can then express the
large distance limit of $\rho^{A}_{\alpha}$ in terms of
$\xi^A_{\delta}$
and the three dimensional Dirac fermion propagator $S_F(x)=\gamma^\mu
x_\mu/(4\pi |x|^2)$ as
\begin{equation}\label{largeDFzeromode}
\rho_\alpha^{{\rm (LD)} ~ A}(\by)=8\pi (S_F(\by-X))^\beta_\alpha
\xi^{A}_\beta\,.
\end{equation}
\paragraph{}
In the four-dimensional theory, the semiclassical dynamics of monopoles
is described by supersymmetric quantum mechanics on the moduli space
\cite{Gauntlett1993}. For a single monopole of mass $M=4\pi|a|/g^2$,
this corresponds to the dynamics of a free non-relativistic
particle moving on $\mathbb{R}^{3}\times S^{1}_{\varphi}$.
These bosonic degrees
of freedom have four free fermionic superpartners. The collective
coordinate Lagrangian takes the form
\begin{equation}\label{MonLQM}
L_{QM}=L_X+L_{\varphi}+L_{\xi}\,,
\end{equation}
where we have $L_X=\frac{M}{2}|\dot{\vec{X}}|^{2}$,
$L_{\varphi}=\frac{1}{2}\frac{M}{|a|^2}(\dot{\varphi})^2$ and
$L_\xi=\frac{M}{2}\dot{{\xi}}^A_{\alpha}\dot{{\xi}_A^{\alpha}}$,
where the dot denotes a time derivative. The
combination $\frac{M}{|a|^2}$ is the moment of inertia of a monopole with
respect to global gauge rotation, $L_{\varphi}$ describes a free particle of
mass $\frac{M}{|a|^2}$ moving along $S_{\varphi}^{1}$ with $\varphi \in
[0,2\pi]$.
\paragraph{}
The quantity of interest here is the large distance behavior of
the four fermion correlation function as defined above. To pass to the
theory on $\mathbb{R}^{3}\times S^{1}$, we Wick rotate the
collective coordinate quantum mechanics described above to a periodic
Euclidean time identified with the $x_{4}$ coordinate introduced
earlier. There are periodic boundary conditions for both bosons and
fermions to preserve supersymmetry.
At leading semiclassical order the fermionic fields in the correlator
are replaced by their values in the monopole background.
The resulting large distance correlation function then takes the following
form:
\begin{eqnarray}
&&\cG_{4}(\by_1, \by_2, \by_3, \by_4)=\int [d\mu] \prod_{A=1}^2
\rho^{{\rm(LD)} ~ A}_1(\by_{2A-1})\rho^{{\rm(LD)} ~
A}_2(\by_{2A})\,,\label{G4}\\
&&\int [d\mu] = \frac{1}{4\pi^{2}}\,\,
\int [d^3 X(x^4)][d\varphi(x^4)][d^4 \xi(x^4)]\cR
\exp\left[-\int^{2\pi R}_0 dx^4 L_{QM}\right]
\exp\left[-\frac{8\pi^2 R |a|}{g^2}+i\th_m\right]\,,\nonumber\\
\label{1measure}
\end{eqnarray}
where superscript ``LD'' on the fermionic zero modes indicates their large
distance behaviors as given in (\ref{largeDFzeromode}). The prefactor
of $1/4\pi^{2}$ arises from the Jacobian for the 
change of variables from bosonic fields to the four bosonic 
collective coordinates and can be traced to the same factor in 
the standard formula \cite{Ber} given as eq.\ (114) in \cite{Dorey1997}.    
The integration measure $[d\mu]$ consists of bosonic $[d^3 X][d\varphi]$
and fermionic $[d^4\xi]$ zero mode measures,
and the one-loop determinant $\cR$ encoding the non-zero mode flucutations,
all weighted by the monopole effective action $\exp[\int^{2\pi R}_0 dx^4
L_{QM}-S_{\rm Mon.}]$ given in (\ref{MonLQM}).
We shall now evaluate various contributions in turns following
\cite{Dorey2000B} and \cite{Kaul}.
\paragraph{}
For the bosonic $\int [d^3 X(x^4)] \exp[\int^{2\pi R}_{0} dx^4 L_{X}]$ and the
fermionic
$\int [d^4\xi(x^4)]\exp[\int^{2\pi R}_{0} dx^4 L_{\xi}]$ zero mode measures,
first we note that $\vec{X}(x^4)$ and $\xi^A_{\alpha}(x^4)$ now need to satisfy
periodic boundary condition
$\vec{X}(x^4)=\vec{X}(x^4+2\pi R)$ and
$\xi^A_{\alpha}(x^4)=\xi^A_{\alpha}(x^4+2\pi R)$.
This implies that for the free Lagragians $L_X$ and $L_\xi$,
the path integrals are dominated by the constant classical paths which, with
slight abuse of notations, we again denote as $\vec{X}$ and
$\xi^A_{\alpha}$.
We can then expand around the classical paths:
\begin{equation}
\vec{X}(x^4)=\vec{X}+\delta
\vec{X}(x^4)\,,~~~\xi_{\alpha}^A(x^4)=\xi_{\alpha}^A+\delta\xi^A_{\alpha}(x^4)\,,
\end{equation}
and decompose the path integrals into
\begin{eqnarray}
\int[d^3 X(x^4)]\exp\left[-\int^{2\pi R}_0 dx^4 L_X\right]
=\int d^3 X \int [d^3\delta \vec{X}(x^4)]\exp\left[-\int^{2\pi R}_0 dx^4
\frac{M}{2}(\delta\dot{\vec{X}}(x^4))^2\right]\,,\label{X3measure}\\
\int[d^4 \xi(x^4)]\exp\left[-\int^{2\pi R}_0 dx^4 L_{\xi}\right]
=\int d^4 \xi \int [d^4\delta \xi (x^4)]\exp\left[-\int^{2\pi R}_0 dx^4
\frac{M}{2}\delta\dot{{\xi}}^A_{\alpha}(x^4)\delta\dot{{\xi}_A^{\alpha}}(x^4)\right]\,.\nonumber
\label{4Fmeasure}\\
\end{eqnarray}
The Gaussian integrals in (\ref{X3measure}) and (\ref{4Fmeasure}) over $\delta
X(x^4)$ and $\delta \xi(x^4)$ can be readily evaluated using standard textbook
results, and we obtain:
\begin{eqnarray}
&&\int [d^3 X(x^4)]\exp\left[-\int^{2\pi R}_0 dx^4 L_X\right]
=\int d^3 X  \left[\sqrt{\frac{M}{2\pi (2\pi R)}}\right]^3\,, \nonumber\\
&&\int [d^4\xi(x^4)]\exp\left[-\int^{2\pi R}_0 dx^4 L_\xi\right]
=\int d^4\xi \left[\sqrt{\frac{M}{2\pi (2\pi
R)}}\right]^{-4}\,.\label{measures}
\end{eqnarray}
Next consider the path integral for $\varphi$, which encodes the motion of the
monopole along $S^{1}_{\varphi}$. The conjugate momentum
$P_{\varphi}=\frac{M}{|a|^2} \dot{\varphi}$ to ${\varphi}$, is
identified with the electric charge and is naturally quantised in
integer units. The corresponding Hamiltonian is
$H_\varphi=\frac{1}{2}\frac{|a|^2}{M} P_{\varphi}^{2}$.
The resulting states in four dimensions carry one unit of magnetic charge
and $P_{e}=n_{e}$ units of electric charge and are naturally
identified as the corresponding BPS dyons.
We can then equate the path integral for  $\int [d\varphi(x^4)]
\exp\left[-\int^{2\pi R}_{0} dx^4 L_\varphi\right]$ with the quantum mechanical
partition function ${\rm Tr}[e^{-(2\pi R) H_\varphi}]$,
where the trace sums over the eigenstates $\sim e^{i\gae \varphi},~\gae \in
\bbZ$ of $H_\varphi$ and can be readily evaluated to give:
\begin{equation}\label{varphiintegral}
\int [d\varphi(x^4)] \exp\left[-\int^{2\pi R}_{0} dx^4
L_\varphi\right]=\sum_{\gae\in \bbZ}
\exp\left[-\frac{1}{2}\frac{|a|^2}{M}\gae^2\right]\,.
\end{equation}
A further phase in the classical action arises from the surface terms
coupling to electric and magnetic charge. Including a bare vacuum
angle $\Theta$  and allowing for the Witten effect which shifts $\gae\to \gae
+\frac{\Theta}{2\pi}$,
the summation in (\ref{varphiintegral}) is replaced by
\begin{equation}
\sum_{\gae\in \bbZ}
\exp\left[-\frac{1}{2}\frac{|a|^2}{M}\left(\gae+\frac{\Theta}{2\pi}\right)^2+
i\left(\gae+\frac{\Theta}{2\pi}\right)\theta_e\right]\,,~~~
M=\frac{4\pi}{g^2}|a|\,.
\end{equation}
We note that this matches the corresponding sum appearing in
the GMN prediction (\ref{gaaApprox}) up to a replacement of the bare
coupling and vacuum angle by their one-loop renormalised counterparts.
\paragraph{}
To complete the semiclassical integration measure, in additon to the zero
modes discussed so far,
it is necessary to include the non-zero mode fluctuations which lead
to a non-cancelling ratio $\cR$ of functional determinants.
Again, we start by reviewing the situation in the four-dimensional
theory where similar fluctuations are taken into account in the calculation
by Kaul \cite{Kaul}
of the one-loop corrections to the monopole mass. In this case,
the {\em spatial} flucutations of the scalars, spinors and ghosts around the
static monopole background are all
described in terms of two operators  $\Delta_{\pm}$ given explicitly
as
\begin{eqnarray}
&&\Delta_+=-D^2_{j}+|a|^2\,,\label{Deltaplus}\\
&&\Delta_-=-D^2_{j}+|a|^2+2 \epsilon_{ijk} \sigma_i F_{jk}^{\rm
Mon.}\,,~~~i,j=1,2,3\label{Deltaminus}\,.
\end{eqnarray}
Here the three dimensional covariant derivative $D_j=\partial_j+ i A_{j}^{\rm
Mon.}$ is with respect to background static monopole
and, as above, $a$ is the VEV of the complex scalar in the massless $U(1)$
vector multiplet.
The one-loop correction to the monopole mass in four dimensions
then involves the
logarithm of the ratio
$[\det(\Delta_+)/\det'(\Delta_-)]^{1/2}$, where the prime indicates
that we have removed the zero mode contribution.
As above, we are interested in the corresponding fluctuations around
the monopole, thought of as a static configuration of finite Euclidean
action on $\mathbb{R}^{3}\times S^{1}$.
In the absence of Wilson line (i.e. for $\theta_{e}=0$), the
corresponding fluctuation operators for our calculation are
\begin{eqnarray}\label{Deltapm}
{\bD}_{\pm}={\Delta}_{\pm}+\left(\frac{\partial}{\partial
x^4} \right)^2\,,
\end{eqnarray}
where the extra derivatives wrt to $x_{4}$ take account of the
Fourier modes of each fluctuation field on $S^{1}$.
We then identify the corresponding one-loop contribution to the path
integral measure as
\begin{equation}\label{Def1loop}
\cR=\left[\frac{\det (\bD_{+})}{\det' (\bD_{-}) }\right]^{1/2}\,.
\end{equation}
\paragraph{}
By translation invariance on $S^{1}$, we
can decompose any eigenfunction of $\bD_{\pm}$ as
$\Phi_{\pm}(\vec{x},x^4)=\phi_{\pm}(\vec{x})f_{\pm}(x^4)$,
where $\phi_{\pm}(\vec{x})$ satisfy
\begin{equation}\label{Defphipm}
{\Delta}_{\pm}\psi_{\pm}(\vec{x})=\lambda_{\pm}^2 \phi_{\pm}(\vec{x})\,,
\end{equation}
while $f_{\pm}(x^4)$ along the compactified circle take the plane-wave form
$f_{\pm}(x^4)\sim e^{i\varpi_{\pm} x^4}$.
In a supersymmetric theory,
there are equal total number of non-zero eigenvalues for both bosonic and
fermionic fields, this naively implies that their contributions cancel
completely and $\cR=1$. However the spectra of $\bD_{\pm}$ contain both
normalizable bound states and continous scattering states, as inherited from
$\Delta_{\pm}$, the precise cancellation requires identical densities of
bosonic and fermionic eigenvalues.
As discovered by \cite{Kaul} this is not the case in the monopole
background. The same effect leads to the non-cancelling
one-loop determinant in the three-dimensional instanton calculation of
\cite{Dorey1997} and we find a similar effect in the
present case of $\bR^3\times S^1$.
\paragraph{}
Using the operator identity $\log {\rm det}({\rm M})={\rm Tr}\log({\rm M})$,
we can rewrite $\cR$ as the following integral expression:
\begin{eqnarray}
\cR &=&(2\pi R )^{-2} \exp\left[\frac{1}{2}\Tr_{\vec{x}} \log [\det
{}_{x^4}
\bD_{+}]-\frac{1}{2}\Tr_{\vec{x}}\log[\det
{}_{x^4}\bD_{-}]\right]\nonumber\\
&=&(2\pi R )^{-2}\exp\left[\frac{1}{2}\int_{|a|}^{\infty} d\lambda
\delta
\rho(\lambda) \log\left[ \cK(\lambda,2\pi R)\right]\right]\,,\label{Int1loop}\\
\cK(\lambda,2\pi R)&=& \det{}_{x^4}\left[\left(\frac{\partial}{\partial
x^4}\right)^2+\lambda^2\right]\,,\label{DefKlambdaR}
\end{eqnarray}
where the overall normalisation constant $(2\pi R)^{-2}$ was introduced so
that $\cR$ goes over to the corresponding
three-dimensional quantity calculated in \cite{Dorey1997} \footnote{To properly
take the three-dimensional limit,
we need to first Poisson re-sum the explicit logarithmic expressions arising in
(\ref{Int1loop}), cf.\ (\ref{cR3}),
and (\ref{dlogD}) in the next section, before setting $R\to 0$.},
\begin{eqnarray}
 \cR^{\rm (3D)} & = &
\left[\frac{\det (\Delta_{+})}{\det' (\Delta_{-})
  }\right]^{1/2}\,\,\,\,=\,\,\, 4 M_W^{2}
\label{3d}
\end{eqnarray}
in the limit $R\rightarrow 0$, $g^{2}\rightarrow 0$ with the
three-dimensional gauge coupling $e^2=\frac{g^{2}}{2\pi R}$ held fixed.
The quantity
$\delta\rho(\lambda)=\rho_{+}(\lambda)-\rho_{-}(\lambda)$
is the difference between densities
of eigenvalues of the operators $\Delta_{+}$ and $\Delta_{-}$.
This quantity was determined using the Callias index theorem in
\cite{Kaul}. In our notation the result of \cite{Kaul} is,
\begin{equation}\label{deltarho}
d\lambda \delta \rho(\lambda)=-\frac{2 |a| d\lambda^2}{\pi
\lambda^2\sqrt{\lambda^2-|a|^2}} \,.
\end{equation}
The remaining kernal $\cK(\lambda,2\pi R)$
is precisely the partition
function of
harmonic oscillator with frequency $\varpi=\lambda$ at inverse temperature
$\beta=2\pi R$.
\begin{equation}\label{cK}
\cK(\lambda, 2\pi R)^{-1}= \frac{\exp[-\pi R \lambda]}{1-
\exp[-2\pi R \lambda]}\,.
\end{equation}
Introducing a non-vanishing Wilson line $\theta_e$ corresponds to
turning on the fourth component of the gauge field. This can be
incorporated in the operators $\bD_{\pm}$ given in (\ref{Deltapm})
by the minimal coupling prescription,
\begin{eqnarray}
\frac{\partial}{\partial x_{4}} & \rightarrow &
\frac{\partial}{\partial x_{4}}\,\,+n_{e}\,\frac{\theta_{e}}{2\pi R}
\nonumber
\end{eqnarray}
which introduces a chemical potential which shifts the oscillator
frequencies to the complex values $\varpi=\lambda+
in_{e}\frac{\theta_e}{2\pi R}$.
The fluctuation modes of each adjoint field include modes with
$n_{e}=\pm 1$ filling out the supermultiplet of the $W^{\pm}$ bosons.
Summing over both contributions we find $\cK=\cK_{+}\cK_{-}$ where
\begin{equation}\label{cKpm}
\cK_{\pm}(\lambda,\theta_e, 2\pi R)^{-1}= \frac{\exp[-\pi R \lambda\pm
i\theta_e/2]}{1-\exp[-2\pi R \lambda \pm i \theta_e]}\,.
\end{equation}
\paragraph{}
Substituting (\ref{cKpm}) and (\ref{deltarho}) into (\ref{Int1loop}), and
explicit change of variable $\lambda=2|a|\cosh t$,
the one-loop determinant $\cR$ is given by:
\begin{eqnarray}\label{cR3}
\log \cR &=& -4R|a|\cosh^{-1} \frac{\Lambda_{\rm
UV}}{|a|}-2\log(2\pi R)\nonumber\\
&+&\frac{2}{\pi }\int^{\infty}_{0} \frac{dt}{\cosh t}\log\left(1-e^{-2\pi
R|a|\cosh t+i \theta_e}\right)
+\frac{2}{\pi }\int^{\infty}_{0} \frac{dt}{\cosh t}\log\left(1-e^{-2\pi
R|a|\cosh t-i \theta_e}\right)\,.\nonumber
\\
\end{eqnarray}
where we have evaluated the integral over the eigenvalues with a UV
cut-off $\Lambda_{\rm UV}$. The UV divergence in the first term is
precisely that encountered in the four-dimensional calculation of
\cite{Kaul}. The divergence is cancelled by the counter-term
which is responsible for coupling constant renormalisation in the vacuum
sector and the net effect is to replace the classical coupling $g^{2}$
appearing in the monopole mass by the
one-loop effective coupling, $g^{2}_{\rm eff}(a)$. Similarly the
chiral anomaly results in the replacement of the classical vacuum
angle $\Theta$ by it effective counterpart $\Theta_{\rm eff}(a)$ as
defined above. The remaining finite terms yield a complicated function of
the dimensionless parameter $|a|R$. However, we recognise the
integral in the second line as precisely the same appearing in the
definition (\ref{sadcD2}) of the quantity $\log \cD(-i)$ in the
semiclassical expansion of the GMN result.
\paragraph{}
Collecting all the pieces and summing over electric charges
$\gae'$, we can extract the
four-fermion vertex in the low-energy effective action
from examining the large distance
behavior of the four fermion correlation function
$\cG_4(\by_1,\by_2,\by_3,\by_4)$.
Substituting (\ref{largeDFzeromode}), (\ref{measures}) and (\ref{cR3}) into
(\ref{G4}) and (\ref{1measure}), the four-fermion correlation function is given
by
\begin{eqnarray}\label{4ptfunction}
\cG_4(\by_1,\by_2,\by_3,\by_4)&=&\frac{2^{13/2}\pi}{R|a|^{1/2}}\cD(-i)\left(\frac{2\pi
R}{g_\eff^2}\right)^{-1/2}
\exp\left[-S_{\rm Mon.}\right]
\sum_{\gae'\in\bbZ}\exp\left[-S_{\varphi}^{(\gae')}\right]
\nonumber\\
&\times &\int d^3 X \epsilon^{\alpha'\beta'}\epsilon^{\gamma'\delta'}
S_F(\by_1-X)_{\alpha\alpha'}S_F(\by_2-X)_{\beta\beta'}S_F(\by_3-X)_{\gamma\gamma'}S_F(\by_4-X)_{\delta\delta'}
\,.\nonumber\\
\end{eqnarray}
Here we have taken into account the one-loop renormalisation effect
discussed earlier in the previous paragraph,
so that the monopole and dyon actions $S_{\rm Mon.}$ and
$S_{\varphi}^{(\gae')}$ are given in terms of the effective parameters,
as:
\begin{eqnarray}
S_{\rm Mon.}&=&\frac{8\pi^2 R}{g_\eff^2}|a|-i\theta_m\,,\\
S_{\varphi}^{(\gae')}&=&\frac{g^2_\eff R
| a|}{4}\left(\gae'+\frac{\Theta_\eff}{2\pi}\right)^2-i\left
(\gae'+\frac{\Theta_\eff}{2\pi}\right)\theta_e\,.
\end{eqnarray}
In (\ref{4ptfunction}), we have also used the relation between
$\cD(-i)$ and $\cR$.
Finally, for consistency, the same renormalisation of the
classical coupling $g^{2}$, which leads to its replacement by the
corresponding effective coupling $g^{2}_{\rm eff}(a)$ in the exponent,
must also be implemented wherever the coupling appears \footnote{Concretely,
this renormalisation corresponds to a divergent contribution to the
instanton measure arising from loop diagrams of perturbation theory in
the monopole background.}. In terms of the low-energy effective action,
the resulting correlator corresponds to the appearance of a
four-fermion interaction term of the form (\ref{transfmn})
\begin{equation}\label{4fermiS}
S_{\rm 4F}=\frac{2^{9/2}\pi}{R|a|^{1/2}} \left(\frac{2\pi
R}{g^2_{\rm eff}(a)}\right)^{7/2}\cD(-i)\exp\left[-S_{\rm Mon}\right]
\sum_{\gae'\in\bbZ}\exp\left[-S_{\varphi}^{(\gae')}\right]\int d^3 x
(\psi\cdot\bar\psi)(\lambda\cdot\bar\lambda)\,,
\end{equation}
we see that
this exactly matches the prediction coming from the integral
equations of \cite{GMN1} given in (\ref{4fermi})!

\section{Interpolating to Three Dimensions}
\paragraph{}
Having matched the predicted action (\ref{4fermi}) and the semiclassical
result (\ref{4fermiS}) for the dyon contributions,
in this section we explain the relation
to the semiclassical instanton result for the
three-di\-men\-sional theory found in \cite{Dorey1997}
which confirms that in the limit $R\to
0$, the hyper-K\"ahler metric on the Coulomb branch is given by Atiyah-Hitchin
manifold \cite{AHmfd}.
To achieve this, we take the metric (\ref{gaainst1}) which
sums over all the electric charges of the dyons $\{\gae'\}$
and Poisson resum it using the standard formula:
\begin{equation}\label{DefPoissonResum}
\sum_{k=-\infty}^{+\infty}f(k)=\sum_{n=-\infty}^{+\infty}\widehat
f(n)\,, \quad
\widehat f(n)=\int_{-\infty}^{+\infty}f(k)\,e^{-2\pi i
n k}dk\,.
\end{equation}
This procedure exchanges the momentum modes along the compact
direction which are identified with the electric charges of dyons
with a corresponding set of winding modes \cite{Dorey2000A}.
This resummation is necessary because the sum over electric charges
appearing in (\ref{gaaApprox})and (\ref{4fermiS}) diverges in the
$R|a| \to 0$ limit.
We can in fact directly perform the Poisson resummation on the
expression (\ref{gaainst1}) for the metric component $g_{a\bar{a}}$.
The relevant
Fourier transform can be evaluated using eq.\ (6.726-4) in
\cite{GandR}\footnote{The required result is obtained by approximating
  the Bessel function in the integrand of eq.\ (6.726-4) by its
  asymptotic form for large arguments.},
allowing us to rewrite the expression as,
\begin{equation}\label{Poissonresum}
\tilde{g}_{a\bar{a}}^{\rm inst.}=
\frac{4\pi}{g_\eff ^4}\sum_{n\in {\mathbb Z}}\frac{|a|^2\cD(-i)}
{|M(n)|^3} \exp\left(-\frac{8\pi^2
R}{g_\eff ^2}|M(n)|+i\Psi(n)\right)\,,
\end{equation}
where we have used the short-hand notation:
\begin{equation}
| M(n)|=\sqrt{|a|^2+\left(\frac{\theta_e}{2\pi R}+\frac{n}{R}\right)^2}\,,\quad
\Psi(n)= \th_m-n\Theta_\eff\,.
\end{equation}
One can also Taylor expand and Poisson re-sum the prefactor
(\ref{sadcD2}) to demonstrate that $\cD(-i)$ satisfies the equation:
\begin{equation}\label{dlogD}
\frac{d \log \cD(-i)}{d(2\pi R|a|)}=2\left(\sum_{n\in \bbZ}\frac{1}{(2\pi
R)|M(n)|}-\frac{1}{\pi}{\rm Arcsinh}\frac{\Lambda_{\rm UV}}{|a|}\right)\,.
\end{equation}
The quantity $M(n)$ appearing in the exponent of (\ref{Poissonresum})
corresponds to the Euclidean action of a ``twisted monopole''
\cite{LeeYi, Dorey1999, Dorey2000A, Dorey2000B}.
These are BPS field configuration in the compactified gauge theory on
${\mathbb{R}}^3\times S^1$ which are obtained by applying a large
gauge transformation of the form
$A_{4}(x)\to A(x)_{4}+\partial \chi(x)$ with $\chi(x_4+2\pi R)=
\chi(x_4)+2n\pi$. As $\theta_e=\int d x^4 A_4$,
the Wilson line undergoes a periodic shift $\theta_e \to
\theta_e+2n\pi$ under this transformation.
These transformations are topologically non-trivial and are
classified by an element
of $\pi_1(S^1)={\mathbb{Z}}$.
This leads to an infinite tower of field configurations for each value of
of the magnetic charge labelled by the winding number $n$. Summing
over these configurations ensures that the metric retains the correct
periodicity in $\th_e$.
\paragraph{}
To compare our prediction with the three-dimensional result, we take
the limit $R\to 0$
while keeping the three-dimensional gauge coupling \footnote{
Note that in \cite{Dorey1997}, the three and four-dimensional
couplings are related via $1/e^2=R/g^2$, which differs from our convention by a
factor of $2\pi$.} fixed: $1/e_\eff^2=2\pi
R/g^2_{\rm eff}$.
Note that $|M(n)| \to \infty$ for $n\neq 0$, and thus only the $n=0$
terms in (\ref{Poissonresum}) and (\ref{dlogD}) survive in this limit.
In the strict three-dimensional limit, $\{{\rm Re}(a),{\rm Im}(a),
\theta_e/2\pi
R\}$ transform as a ${\bf 3}$ under global $SU(2)_N$ symmetry, hence we can
rotate
into $\theta_e=0$ vacuum, and (\ref{dlogD}) integrates into $\cD(-i)=(4\pi R
| a|)^2$.
Using this symmetry, (\ref{expRiemanntensor}), and the normalization factors
given in
(\ref{normalization})-(\ref{2fermions}), we can then deduce that the following
four-fer\-mions vertex is generated in the low-energy effective action:
\begin{equation}\label{3D4Fvertex}
S_{4F}=\frac{2^{7}\pi^{3} M_W}{e_\eff^8}
\exp\left(-\frac{4\pi}{e_\eff^2}M_W+i\theta_m\right)\int
d^3x\,(\psi \cdot \bar{\psi})(\lambda \cdot \bar{\lambda})\,,
\end{equation}
where $M_W=\sqrt{|a|^2+(\theta_e/(2\pi R))^2}$. Comparing with eqs.\ (29, 34)
in
\cite{Dorey1997}, we obtained a perfect match with the four-fermion vertex
generated from the three-dimensional one instanton semiclassical computation,
which was also in agreement with the prediction coming from the exact
Atiyah-Hitchin metric given in eq.\ (54) of \cite{Dorey1997}.

\subsection*{Acknowledgements}
HYC is supported in part by NSF CAREER Award No. PHY-0348093, DOE grant
DE-FG-02-95ER40896, a Research Innovation Award and a Cottrell Scholar Award
from Research Corporation, and a Vilas Associate Award from the University of
Wisconsin.
KP is supported by a research studentship from Trinity College, Cambridge.

\end{document}